\documentstyle[12pt]{article}
\def\g{\gamma}

\def\th{\theta}

\def\y{\psi}
\def\by{\bar{\psi}}
\def\r{\rho}

\def\f{\phi}

\def\bg{\begin{eqnarray}}
\def\ed{\end{eqnarray}}

\def\la{\langle}
\def\vp{\varphi}
\def\baF{\tilde{F}}
\def\ra{\rangle}
\def\fd{\phi ^{\dagger}}

\def\lm{\lambda}
\def\z{${\bf Z_2}$}

\def\D{\not\!\! D}

\begin{document}
\begin{flushright}
SFU-Preprint-92-7 \\
\end{flushright}
\vskip.20in
\begin{center}
{\Large\bf On the Anomalous Discrete Symmetry}
\\*[.35IN]
{\bf Zheng Huang} \\
Department of Physics, Simon Fraser University\\
 Burnaby, B.C., Canada V5A 1S6\\
\end{center}
\vskip .50in

\begin{abstract}
We examine an interesting scenario to solve the domain wall problem
recently suggested by Preskill, Trivedi, Wilczek and Wise. The effective
potential is calculated in the presence of the QCD axial anomaly. It is
shown that some discrete symmetries such as CP and \z can be
anomalous due to a so-called $K$-term induced by instantons. We point out
that \z domain-wall problem in the two-doublet standard model
can be resolved by two types of solutions: the CP-conserving
one and the CP-breaking one. In the first case, there exist two
\z -related local minima whose energy splitting is provided by
the instanton effect. In the second case, there is only one
unique vacuum so that the domain walls do not form at all. The
consequences of this new source of CP violation are discussed
and shown to be well within the
experimental limits in weak interactions.
\end{abstract}
\vskip .1in
PACS numbers: 12.15.Cc, 11.30.Er, 11.30.Qc.\\
\vskip .1in
\today
\pagebreak

Recently, Preskill, Trivedi, Wilczek and Wise (PTWW) \cite{ptww} have reported
an
interesting scenario to solve the cosmological domain wall problem
associated with spontaneously broken discrete symmetry. They have pointed
out that because some discrete symmetry can be anomalous due to the QCD
axial anomaly and instantons, a non-perturbative communication between
the Higgs sector and the QCD sector leads to a tiny but cosmologically
significant splitting of the vacuum degeneracy. Incorporating PTWW's idea,
Krauss and Rey \cite{kr} have shown that certain models of spontaneous CP
violation can in principle avoid the domain wall problem provided that CP is
slightly broken by $\th _{\mbox{QCD}}$ in strong interactions. In this letter,
we
examine the idea by computing the effective potential for Higgs bosons in the
presence of QCD chiral anomaly. We show that the instanton dynamics for light
quarks does break \z symmetry of the two-doublet standard model. However, it
may
also lead to a spontaneous  CP symmetry breaking.

To illustrate how the anomalous discrete symmetry arises, let us first consider
a simplest model with  spontaneous CP violation. The prototype of this model
was first considered by T.\ D.\ Lee \cite{lee} where the Higgs field $\vp$
belongs to a real representation,
\bg
{\cal L}_0=\frac{1}{2}(\partial \vp)^2-\lm ^2 (\vp ^2-\eta ^2)^2+
\by (\not \!\partial +m-if\g _5\vp )\y .\label{1}
\ed
The minimum of the potential corresponds to $\la \vp \ra =\pm \eta$ and CP
symmetry is spontaneously broken. It was first pointed out by Kobsarev, Okun
and Zeldovich (KOZ) that the degeneracy of CP conjugate vacua $\la \vp \ra
=\eta$
and $\la \vp\ra =-\eta $ results in a serious domain wall problem in cosmology
\cite{koz}. However, the situation is quite different if the fermion field $\y$
suffers from non-abelian gauge interactions. In that case, (\ref{1}) can be
extended to include, for example,
color interactions ($\vp $ is of course colorless)
\bg
{\cal L}=\frac{1}{2}(\partial \vp)^2-\lm ^2 (\vp ^2-\eta ^2)^2+
\by (\not \!\! D +m-if\g _5\vp )\y -\frac{1}{4}F^2-i\th F\baF \label{2}
\ed
where $F\baF =\frac{g^2}{32\pi
^2}\epsilon_{\mu\nu\r\sigma}F^{\mu\nu}F^{\r\sigma}$.
Though CP is explicitly broken by the $\th$-term if $\th\neq
0,\pi$, the domain wall problem persists at the tree level
because the vacua $\la \vp\ra=\pm \eta$ are still in
degeneracy. However, we show that the degeneracy of the vacua will be lifted by
taking into account the chiral anomaly or
the instanton effect.

The effective action of the Higgs field is calculated as
\bg
{\bf Z}={\int {\cal D}(\vp)e^{-S_0[\vp ]}{\int {\cal D}(A,\by ,\y )e^{-S[
\by ,\by ;A;\vp ]}}}={\int {\cal D}{\vp}e^{-S_{eff}[\vp ]}} \label{3}
\ed
where
\bg
S_{eff}[\vp ]=S_0[\vp]+\Delta S[\vp] \label{4}
\ed
and the quantum correction is given
\bg
\Delta S[\vp ]=-\ln {\int {\cal D}(A;\by ,\y )e^{-S[\by ,\by ;A;\vp ]}}\equiv
-\ln {\bf \tilde{Z}}.\label{5}
\ed
The calculation of ${\bf \tilde{Z}}[\vp ]$ in the instanton field follows
the standard semiclassical approximation method as illustrated in, e.\ g.\ ,
Ref.\cite{thooft}
\bg
{\bf \tilde{Z}}[\vp ]=\sum _\nu \int {\cal D}[A_{cl}]_\nu e^{-S[A_{cl}]}{\det}
^{-1/2}
M_A{\det} M_{\y}{\det} M_{gh} \label{6}
\ed
where
\bg
M_A & = & -D^2-2F \nonumber \\
M_{gh}& = & -D^2 \label{7}\\
M_\y & = & \not \!\! D+m-if\g _5\vp \nonumber
\ed
and $\nu$ stands for the winding number of the non-trivial topological
gauge configuration. If the effective potential is of concern, we can take
$\vp$ in $M_\y$ as a constant field.
The new physics comes from the zero modes of the fermion determinant in the
instanton
field $A_{cl}$. We factorize ${\det} M_\y$ as follows
\bg
{\det} M_\y ={\det} ^{(0)}M_\y {\det} 'M_\y \label{8}
\ed
where ``${\det} ^{(0)}$'' denotes contributions from the subspace of zero modes
of $\not \!\! D$. According to the index theorem \cite{as}, $\not\!\! D$ has a
zero
mode with chirality $-1$ ($\g _5=-1$) in a single instanton field \cite{brown}.
Thus we have
\bg
{\det} ^{(0)}M_\y =m+if\vp . \label{9}
\ed
The prime in ${\det} 'M_\y$ reminds us of excluding zero modes from the
eigenvalue
product. Since $[\D ,\g _5]\neq 0$, $M_\y$ cannot be diagonalized in the basis
of
eigenvectors of $\D$. The non-vanishing eigenvalues of $\D$ always appear in
pair, i.\ e.\ if $\D \vp _n=\lm _n\vp _n$ where $\lm _n\neq 0$, then
$\D \g _5\vp _n=-\g _5\D \vp _n=-\lm _n\g _5\vp _n$, namely both $\lm _n$ and
$-\lm _n$ are eigenvalues of $\D$. In addition, $\g _5$ takes $\vp _n$ to
$\vp _{-n}$. Therefore
\bg
{\det} 'M_\y &=&\prod _{\lm _n>0}{\det}
\left( {\matrix{{i\lambda _n+m}&{-if\varphi }\cr
{-if\varphi }&{-i\lambda _n+m}\cr
}} \right)=\prod _{\lm _n>0}(\lm _n^2+m^2+f^2\vp ^2) \nonumber\\
 &=& {\det} '^{1/2}(-\D^2+m^2+f^2\vp ^2), \label{10}
\ed
i.\ e.\ ${\det}'M_\y$ is a function of $\vp ^2$ which does not break the
discrete symmetry.
It is to be emphasized that the above analysis does not depend on the detail
of the instanton dynamics. It is the result of using the index theorem, which
represents the general feature of the chiral anomaly in a gauge theory.

Though we could proceed to analyze in general the effective potential
based on Eqs.\ (\ref{9}) and (\ref{10}), we still would like to obtain the
concrete form of $V_{eff}$ in the dilute gas approximation (DGA)
\cite{callanet}. In the DGA,
\bg
{\bf \tilde{Z}}[\vp ]=\det (-\partial ^2+m^2+f^2\vp ^2)\exp
({\bf \tilde{Z}_+}+{\bf \tilde{Z}_-}) \label{11}
\ed
where
\bg
{\bf \tilde{Z}_+}[\vp ]&=& VKe^{i\th}(m+f\vp ) \nonumber \\
{\bf \tilde{Z}_-}[\vp ]&=& VKe^{-i\th}(m-f\vp ) \label{12}
\ed
and
\bg
K=1.34C_{N_c}\int \frac{d\r}{\r ^4}\left(\frac{8\pi ^2}{g^2(\r )}\right)^{2N_c}
e^{-\frac{8\pi ^2}{g^2(\r )}}. \label{13}
\ed
$\r$ is the instanton density, $C_{N_c}=\frac{N_c^2-1}{2N_c}$, $N_c$
is the number of colors. In deriving (\ref{12}), we have assumed that
$m+f\la \vp\ra$ is small compared to $\Lambda _{\mbox{QCD}}$. Noticing that
$\ln \det (-\partial ^2+m^2+f^2\vp ^2)$ contains terms which can be
absorbed into the tree level lagrangian by redefining $\lm ^2$ and
$\eta$, we obtain the following effective potential (strictly
speaking in the large $N_c$ limit)
\bg
V_{eff}=\lm ^2 (\vp ^2-\eta ^2)^2+Ke^{i\th} (m+f\vp )+
Ke^{-i\th}(m-f\vp ). \label{14}
\ed
Clearly, the last two terms (we shall call them the $K$-term)
$explicitly$ break CP symmetry when $\th\neq 0$, for they are not
invariant under $T\vp T^{-1}=-\vp$. The split in the energy density
between the CP conjugate vacua $\la \vp\ra =\eta$ and
$\la \vp\ra=-\eta$ is given
\bg
\Delta E_{vac}=\left| V_{eff}(\eta )-V_{eff}(\eta )\right|=4Kf\sin\th |\la \vp
\ra |. \label{15}
\ed
Therefore, domain walls created at the scale $\la \vp\ra$ will feel
an energy difference between the two sides of the wall. The false
vacuum at some space point will begin to decay towards the true vacuum.

Another perhaps more interesting example to observe the anomalous
discrete symmetry is to consider the two Higgs doublets model, which
is the simplest allowed extension of the standard model. To achieve
the natural  neutral flavor conservation (NFC) at the tree level, we
impose Glashow-Weinberg's \z discrete symmetry: $\f _1$ couples with
the charge $\frac{2}{3}$ quarks ($U_R$) and $\f _2$ couples to the
charge $-\frac{1}{3}$ quarks ($D_R$), i.\ e.\ , for  example,
\bg
\f _1\rightarrow \f _1, \quad \f _2\rightarrow -\f _2;
\quad\quad U_R\rightarrow U_R, \quad D_R\rightarrow -D_R. \label{16}
\ed
The most general, renormalizable Higgs potential and Yukawa interactions which
respect (\ref{16}) read
\bg
V_0(\f _1,\f _2)&=&-m_1^2\fd _1\f _1-m_2^2\fd _2\f _2+a_{11}(\fd _1\f _1)^2
+a_{22}(\fd _2\f _2)^2     \label{17}\\
& & +a_{12}(\fd _1\f _1)(\fd _2\f _2)+b_{12}(\fd _1\f _2)(\fd _2\f _1)
+\lm [(\fd _1\f _2)^2 +(\fd _2\f _1)^2] \nonumber
\ed
and
\bg
{\cal L}_Y=\bar{Q}_Lf_UU_R\f _1+\bar{Q}_Lf_DD_R\tilde{\f}_2+\mbox{h.c.}
\label{18}
\ed
Here $f_U$ and $f_D$ are $3\times 3$ Yukawa coupling matrices in flavor space,
$\tilde{\f}_2=i\sigma _2\f _2^*$. The hermicity of $V_0$ requires all
coefficients in (\ref{17}) are real. We shall examine the spontaneous CP
violation (SCPV) in this model, thus we first choose $f_U$ and $f_D$ to be real
and
$\th _{\mbox{QCD}}=0$ in the QCD sector. When $\f _1$ and $\f _2$ acquire
VEV's, \z
symmetry (\ref{16}) is spontaneously broken, which poses dangers for
cosmology. PTWW have argued that when the non-perturbative QCD effect turns
on, it breaks \z symmetry and solves the domain wall problem.

To see explicitly how PTWW's idea works, we attempt to compute the $K$-term in
the effective potential following the same procedure as we did in the previous
model.
I will first consider one generation of light quarks consisting
of $u$ and $d$ ($m_u, m_d\ll \Lambda _{\mbox{QCD}}$) to simplify the
problem.  The Higgs
coupling to light quarks can be rewritten in a form
\bg
{\cal L}_m=
\left( {\matrix{{\bar u_L}&{\bar d_L}\cr
}} \right)H\left( {\matrix{{u_R}\cr
{d_R}\cr
}} \right)+\left( {\matrix{{\bar u_R}&{\bar d_R}\cr
}} \right)H^\dagger\left( {\matrix{{u_L}\cr
{d_L}\cr
}} \right) \label{19}
\ed
where
\bg
H\equiv \left( {\matrix{{f_d\phi _2^0*}&{f_u\phi _1^+}\cr
{-f_d\phi _2^-}&{f_u\phi _1^0}\cr
}} \right).\label{20}
\ed
Thus it is easy to identify
\bg
M_\y =\D + \frac{1}{2} (H+H^{\dagger})+\frac{1}{2}(H-H^{\dagger})\g _5
\label{21}
\ed
where $\det M_\y$ runs over color, spinor as well as flavor indices,
\bg
{\det}^{(0)}M_\y =\left\{ \begin{array}{ll}
\det H^{\dagger}=f_uf_d\fd _1\f _2 & \mbox{for a single instanton}\\
\det H=f_uf_d\fd _2\f _1 & \mbox{for a single anti-instanton}
\end{array}
\right. \label{22} \ed
and
\bg
{\det}'M_\y &=&{\det}'^{1/2}(-\D ^2+HH^{\dagger})\label{23}\\
 &=& \prod _{\lm _n>0}[\lm _n^4+\lm _n^2(f_u^2\fd _1\f _1+f_d^2\fd _2\f _2)
+f_u^2f_d^2(\fd _1\f _2)(\fd _2\f _1)].\nonumber
\ed
It is clear that ${\det}'M_\y$ can be absorbed into $V_0(\f _1,\f _2)$ in
(\ref{17}) but ${\det}^{(0)}M_\y$ constitutes the so-called the $K$-term which
breaks \z symmetry.  The effective potential reads
\bg
V_{eff}(\f _1,\f _2)=V_0(\f _1,\f _2)+Kf_uf_d(\fd _1\f _2+\fd _2\f _1)
\label{24}
\ed
where
\bg
K=(1.34)^2C_{N_c}\int \frac{d\r}{\r ^3}\left(\frac{8\pi ^2}{g^2(\r
)}\right)^{2N_c}
\exp \left(-\frac{8\pi ^2}{g^2(\r )}\right). \label{25}
\ed
$K$ is of dimension 2.

When $\lm <0$ ($\lm$ is the coefficient of $[(\fd _{1}\f
_{2})^{2}+(\fd _{2}\f _{1})^{2}]$ in (\ref{17})), it can be
readily shown that the \z -related $(v_{1},v_{2})$ and
$(v_{1},-v_{2})$ (where $v_{1}$ and $v_{2}$ are real) are
local minima of $V_{eff}(\f_{1},\f_{2})$. However, they are not
degenerate because of the $K$-term. The difference in the energy density
 between these two vacua
$(v_1,v_2)$ and $(v_1,-v_2)$ is given
\bg
\Delta
E_{vac}=\left|V_{eff}(v_{1},v_{2})-V_{eff}(v_{1},-v_{2})\right|\simeq
4Kf_uf_dv_1v_2=4Km_um_d. \label{26}
\ed
$K$ is the vacuum-to-vacuum amplitude in the instanton field. It is also the
amplitude of the axial $U(1)$ symmetry breaking in QCD needed to solve the
$U(1)$ problem. It has been estimated in \cite{huang} in connection with the
$U(1)$ particle mass
\bg
K\sim (m_\eta ^2-m_\pi ^2). \label{27}
\ed
Thus $\Delta E_{vac}\simeq 10^{-4}\sim 10^{-5} \mbox{GeV}^4$, which is tiny
but significant enough to solve the domain wall problem associated with \z
symmetry \cite{ptww}. When $\lm >0$, none of $(v_1,v_2)$ and $(v_1,-v_2)$
are minima. In fact, they are both local maxima of $V_{eff}$. The
true vacuum, denoted by $(v_{1},v_{2}e^{i\alpha})$, which minimizes the
effective potential acquires a
non-trivial phase $\alpha$ ($\alpha\neq 0,\pi$). The domain
wall problem associated with \z is automatically resolved since
$(v_{1},-v_{2}e^{i\alpha})$ is no longer the minimum of the
potential.

However, what interests us is that the existence of the
relative phase between $\la \vp _{1}\ra$ and $\la \vp _{2}\ra$
breaks CP symmetry spontaneously. To see this, we calculate the
$\alpha$-dependent terms in the effective potential
\bg
V_{eff}(\alpha )=2\lm v_{1}^{2}v_{2}^{2}\cos 2\alpha +
2Kf_{u}f_{d}v_{1}v_{2}\cos \alpha . \label{28}
\ed
By minimizing $V_{eff}(\alpha )$ with respect to $\alpha$ one
obtains
\bg
\cos \alpha =-\frac{Km_{u}m_{d}}{4\lm v_{1}^{2}v_{2}^{2}},
\label{29}
\ed
which is about $10^{-14}$ if $v_{1}$ and $v_{2}$ are taken to
be the electroweak scale. Therefore, Triggered by strong
interactions, CP  is spontaneously broken at the electroweak
scale in the two-doublet model. It is well known that \z
symmetry in the two-doublet model actually forbids the
spontaneous CP violation. However, when \z is explicitly broken
by instanton effects, the SCPV is allowed but with a
dynamically determined magnitude.

Does this new source of CP violation lead to any observable
effects in electroweak interactions? Obviously, the phases of
quark masses and Yukawa couplings originating from SCPV can be
rotated away by making appropriate hypercharge transformation.
Thus the CP-breaking Cabbibo-Kobayashi-Maskawa (CKM) matrix
does not arise in this model. The CP nonconservation is
entirely due to neutral Higgs boson exchanges, i.\ e.\ through
the mixing between scalar fields and pseudoscalar fields  while
the mixing probability is proportional to $\sin \alpha\cos
\alpha$ which is about $10^{-14}$. All CP-violating processes
are to be suppressed by this factor. Its contribution to
$K_{L}\rightarrow 2\pi$ can be neglected since this process
involves charged flavor changing. The electric dipole moment of
neutron (NEDM) will be receiving suppression factors, a
$10^{-8}$ from Higgs propagators if Higgs bosons are of
$100$GeV and a $10^{-14}$ form the mixings. Thus the NEDM is
estimated to be $10^{-34}e\cdot cm$, which is not practically
detectable. It is also not sufficient to generate the
electroweak baryogenesis based on the weak phase transition
since  the instanton effect is greatly suppressed at
temperature characteristic of the weak scale \cite{linde}. Even
though there are several ways of enhancing the CP violating
effects by, for example, allowing a large ratio of $v_{1}$
to $v_{2}$ or having nearly degenerate masses for Higgs bosons,
it would seem unnatural to yield any sizable observations.

The evaluation of the relative phase between two vacuum
expectation values can be readily generalized to including any
number of quark generations and explicit CP violation in a
manner of KM mechanism without resorting to the instanton
computations. In general, the Yukawa coupling matrices $f_{U}$
and $f_{D}$ can be complex. The phases of their determinants
can be rotated away by redefining the right-handed quark fields
while in the meantime chaning  $\th _{\mbox{QCD}}$, the coefficient of the QCD
$\th$-term. We parametrize $\f _{1}$ and
$\f _{2}$ in terms of  their phase fields $\alpha _{1}(x)$ and $\alpha
_{2}(x)$ as
\bg
\f _{1}\longrightarrow v_{1}e^{i\alpha _{1}(x)}
\quad\quad\quad;\quad\quad\quad \f _{2}\longrightarrow
v_{2}e^{i\alpha _{2}(x)} \label{30}
\ed
and denote the relative phase field by $\alpha (x)\equiv \alpha
_{1}(x)-\alpha _{2}(x)$. The $\alpha _{1}$- and $\alpha
_{2}$-dependence of the Yukawa couplings can be removed by
making the $local$ chiral rotations. Because of the chiral
anomaly, the $\th$-term becomes
\bg
\left (\th _{\mbox{QCD}}+n_{G}\alpha (x) \right)F\baF \label{31}
\ed
where $n_{G}$ is the number of the quark generations. The
effective potential for $\alpha (x)$ is calculated \cite{shifman}
\bg
V_{eff}=-\la\la \nu ^{2}\ra\ra _{\mbox{QCD}}\cos \left( \th
_{\mbox{QCD}}+n_{G}\alpha\right) +2\lm v_{1}^{2}v_{2}^{2}\cos 2\alpha
\label{32}
\ed
where the topological susceptibility $\la\la \nu\ra\ra _{\mbox{QCD}}$
is defined
\bg
\la\la\nu ^{2}\ra\ra _{\mbox{QCD}}=\int d^{4}x \la \mbox{T}iF\baF (x)\;
iF\baF (0)\ra . \label{33}
\ed
By minimizing (\ref{32}) one obtains
\bg
\sin 2\alpha =\frac{n_{G}\la\la\nu ^{2}\ra\ra _{\mbox{QCD}}\sin
\bar{\th}}{4\lm v_{1}^{2}v_{2}^{2}} \label{34}
\ed
where $\bar{\th}=\th _{\mbox{QCD}}+n_{G}\alpha$.

It is then clear from (\ref{34}) that when $\th _{\mbox{QCD}}\neq 0$,
$\alpha =0,\pi$ are not extremes of $V_{eff}(\alpha )$, i.\ e.\
both $(v_{1},v_{2})$ and $(v_{1},-v_{2})$ are not local minima
of the effective potential. The \z domain-wall problem is
resolved by admitting a CP-violating solution
$(v_{1},v_{2}e^{i\alpha})$. In this case, the weak CP violation
is further suppressed by the requirement that $\sin
\bar{\th}<10^{-9}$ from the strong CP violation. When $\th
_{\mbox{QCD}}=0$ and $\lm <0$, (\ref{34}) gives the solution provided
by PTWW, i.\ e.\ both $(v_{1},v_{2})$ and $(v_{1},-v_{2})$ are
local minima whose energy splitting is caused by instanton
effects. When $\th _{\mbox{QCD}}=0$ and $\lm >0$, (\ref{34}) allows a
non-trivial CP-breaking solution $\alpha \neq 0,\pi$. In this
case \z domain walls do not form but domain walls associated
with SCPV ($\alpha$ and $-\alpha$) will begin to form.

I would like to thank K.S.\ Viswanathan and D.D.\ Wu for useful
discussions.
\pagebreak

\end{document}